\documentclass[prd,nofootinbib,superscriptaddress,twocolumn,preprintnumbers]{revtex4}
\usepackage{graphicx}
\usepackage[english]{babel}
\usepackage{amsmath,amssymb}

\usepackage{color}

\newcommand{\fref}[1]{Fig.\ref{#1}}

\newcommand{\eref}[1]{Eq.(\ref{#1})}
\newcommand{\erefs}[1]{Eqs.(\ref{#1})}
\newcommand{\reff}[1]{(\ref{#1})}
\newcommand{\citer}[1]{Ref.\cite{#1}}
\newcommand{\citers}[1]{Refs.\cite{#1}}

\newcommand{\p}{\partial}
\newcommand{\ie}{\emph{i.e., }}

\newcommand{\prm}{\mathrm{Pm}}
\newcommand{\mff}{\psi}
\renewcommand{\vec}{\boldsymbol}
\begin{document}
\title{Behavior of thin disk crystalline morphology\\in the presence of corrections to ideal magnetohydrodynamics}
\author{Giovanni Montani}
\affiliation{ENEA, Fusion and Nuclear Safety Department, \\ C.R. Frascati, Via E. Fermi 45, 00044, Frascati, Roma, Italy}
\affiliation{Department of Physics, ``Sapienza'' University of Rome, \\ Piazzale Aldo Moro 5, 00185, Roma, Italy}
\author{Mariachiara Rizzo}
\affiliation{Department of Physics, ``Sapienza'' University of Rome, \\ Piazzale Aldo Moro 5, 00185, Roma, Italy}
\author{Nakia Carlevaro}
\affiliation{ENEA, Fusion and Nuclear Safety Department, \\ C.R. Frascati, Via E. Fermi 45, 00044, Frascati, Roma, Italy}
\affiliation{L.T. Calcoli, Via Bergamo 60, 23807, Merate, Lecco, Italy}

\begin{abstract}
We analyze an axisymmetric magnetohydrodynamics configuration, describing the morphology of a purely differentially rotating thin plasma disk, in which linear and non-linear perturbations are triggered associated with microscopic magnetic structures. We study the evolution of the non-stationary correction in the limit in which the co-rotation condition (\ie the dependence of the disk angular velocity on the magnetic flux function) is preserved and the poloidal velocity components are neglected. The main feature we address here is the influence of ideal (finite electron inertia) and collisional (resistivity, viscosity, and thermal conductivity) effects on the behavior of the flux function perturbation and of the associated small-scale modifications in the disk. We analyze two different regimes in which resistivity or viscosity dominates and study the corresponding linear and non-linear behaviors of the perturbation evolution, \ie when the backreaction magnetic field is negligible or comparable to the background one, respectively. We demonstrate that when resistivity dominates, a radial oscillating morphology (crystalline structure) emerges and it turns out to be damped in time, in both the linear and non-linear regimes, but in such a way that the resulting transient can be implemented in the description of relevant astrophysical processes, for instance, associated with jet formation or cataclysmic variables. When the viscosity effect dominates the dynamics, only the non-linear regime is available and a very fast instability is triggered.
\end{abstract}
%\pacs{97.10.Gz, 95.30.Qd}

\maketitle

\section{Introduction}
One of the most intriguing open questions in theoretical astrophysics is the mechanism underlying the transport processes of accretion plasma disks around a compact object \cite{Bis2001,Bal98}. The most commonly accepted idea essentially relies on the original Shakura proposal \cite{S73} (see also \citer{SS74}), which consists of postulating an effective plasma viscosity, able to account for the angular momentum transport. Clearly, such a dissipation effect can not originate from the kinetic properties of the plasma, which is essentially ideal for most of the plasmas in accreting astrophysical systems. Instead, the required viscosity arises from the turbulent plasma behavior. In fact, it is well known that convection disk instability saturates into a turbulent regime able to enhance the plasma effective shear viscosity \cite{Bal98,lp80,rpl88}. As a consequence, under a suitable averaging procedure (mainly based on a full azimuthal average and local radial and vertical ones), the dynamics resembles a laminar flow in the presence of effective viscosity; the non-ideal terms come from the correlation function of the turbulent velocity field components. Actually, the standard model of accretion disks (\ie the $\alpha$ disk model) relies on the idea that all the supersonic fluctuations are suppressed as time goes by and the correlation function of radial and azimuthal velocity components is, on average, estimated by $\alpha v_s^2$ (where $\alpha$ is a parameter less than unity and $v_s$ denotes the sound velocity of the plasma disk).

The basic plasma instability able to generate, via its saturation, the requested turbulence can be identified in the so-called magnetorotational instability (MRI) \cite{Vel59,Chan60,Bal91} (see also \citer{Bal95} and, for a global approach, \citer{PAPA92}). MRI is due to the coupling of the Alfv\'en modes to the differential rotation of the disk. This instability exists only in weakly magnetized plasmas, as many disk regions turn out to be, far enough from the central object and therefore it is a reliable scenario for the implementation of MRI as the trigger for the turbulent regimes, able to account for the angular momentum transport across the disk via an effective shear viscosity coefficient.

However, introducing a magnetic field in the problem requires that also the generalized Ohm law must be satisfied in the plasma and since the currents induced in the disk are in general very small, this implies an effective large value of the resistivity coefficient. This is also known as anomalous resistivity and it calls for a convincing explanation, especially in those astrophysical systems, like x-ray binaries, for which the mass accretion rate is particularly large (see the discussion presented in \citer{MC12}).

An alternative perspective has been traced in \citers{1,2} (see also \citer{11}), where the possibility of an oscillating radial behavior of the backreaction (crystalline magnetic micro-structure) was investigated, and then extended from a local to a global picture in \citer{Ben2011}. Despite such a reformulation of the local plasma equilibrium still being far from an alternative reliable accretion model, it nonetheless appears as a valuable cross over from laboratory plasma physics and it has two main advantages: (i) The short characteristic spatial scale of the magnetic field structures allows one to deal with larger values of the current densities so that the anomalous values of resistivity can be avoided and (ii) the magnetic field, having no diffusive profiles as in the standard resistive picture, can increase its values in some regions of the disk, thus offering a possible paradigm for the generation of collimated jets \cite{CM10,TCM13}. However, in \citer{Petitta2013} it was shown how the magnetic micro-structures can be damped by viscous-resistive effects, acquiring the morphology of short transients in many contexts of astrophysical interest.

The present study generalizes the analysis in \citer{Petitta2013} by including, in addition to viscosity and resistivity, the effect of a finite electron inertia (an ideal contribution expected to be important for low values of the plasma parameter $\beta$). Here we analyze the evolution of magnetic micro-structures in both the linear and non-linear regimes, \ie when the backreaction magnetic field is small or comparable to the background one, respectively. We consider, as in \citer{Petitta2013}, a purely differentially rotating background, embedded in a poloidal magnetic field and we assume the validity of a co-rotation condition, \ie the disk angular velocity is, at any order of approximation, expressed via the magnetic flux function. The plasma disk configuration is considered thin, according to the most common disk morphology \cite{Bis2001}, and due to the small spatial perturbation scale, we deal with a local model for which a fiducial value of the distance from the central compact object is considered.

The present analysis has two main merits. (i) We demonstrate that, in the presence of finite electron inertia, the damped crystalline profile outlined in \citer{Petitta2013} still survives, but now the magnetic Prandtl number (MPN) is no longer strictly constrained to be equal to one. The model is now applicable, in principle, for any value of such a parameter between $0$ and $1$. Actually, as discussed in \citers{balhenri08,Bal98}, the $\alpha$ disk model is associated with very small values of the MPN except for black-hole and neutron-star accretion disks for which it can be larger, with non-trivial implications concerning the turbulence features of MRI saturation. Furthermore, this range of the MPN has the important consequence that the life-time of the micro-structures is significantly enhanced. (ii) Furthermore, we show that for a MPN greater than one, a non-linear instability exists, able to enhance the radial profile of the perturbations, so triggering the onset of a new physical regime of the disk. In other words, we find a bifurcation in the perturbation behavior: As far as they remain sufficiently small in amplitude, the disk is characterized by a damped radial corrugation, but if the plasma backreaction is strong, depending on whether the viscous or resistive effects dominate, the profile can acquire a new growing behavior (non-linear instability) or still follow the damped regime, respectively. According to the paradigm inferred in \citers{CM10,TCM13} for the jet generation from the crystalline profile of the perturbed accreting plasma, we are led to consider the present non-linear growing behavior of the disk corrugation (in the presence of finite electron inertia) as an interesting mechanism to trigger the formation of collimated energetic structures in the disk morphology.

\section{Fundamental equations}
The analyzed system is a geometrically thin, non-self-gravitating disk of plasma in differential rotation around a central stellar object. We adopt cylindrical coordinates $(r,\, \phi,\, z)$, where $z$ is the axis of symmetry. The electric and magnetic fields $\vec{E}$ and $\vec{B}$, respectively, and the current density field $\vec{J}$ can be expressed via the magnetic flux function $\mff$, defined as
\begin{align}
\mff = \int_0^r 2 \pi r' B(r',z) dr'\;,
\end{align}
in the form
\begin{align}
\vec{B} = -\frac{1}{r} \partial_z \mff \hat{e_r} +\frac{1}{r} \partial_r \mff \hat{e_z}\label{B}\;,\\
\vec{E}= \nabla \Phi - \frac{1}{c} \partial_t \vec{A}\;,\\
\vec{A}= \frac{\mff}{r} {\hat{e}}_{\phi}\;,\\
\vec{J}= -\frac{c}{4 \pi} \nabla \times \vec{B}\;,
\end{align}
where $\vec A$ is the vector potential (such that $ \vec{B} =\nabla \times \vec{A}$), while $\Phi$ denotes the electric scalar potential. We adopt a perturbation scheme, in which we split all the physical quantities into two parts: a background contribution (denoted by the subscript $0$) and a perturbative term (denoted by the subscript $1$). In particular, we write
\begin{equation}
\mff= \mff_0 (R_0) + \mff_1 (R_0, r- R_0, z)\;,
\label{perturbazione}
\end{equation}
where $|\mff_1|\ll|\mff_0|$. Here we face a local analysis by setting $R_0$ as the fiducial distance from the center of the stellar object, around which the problem is developed. While $|\mff_1|\ll |\mff_0|$, the correction $\mff_1$ is assumed to be a small-scale varying function, \ie its derivatives can be of the same order as or greater than the background one, and thus its contribution to the magnetic field can be relevant.

The main point of this study is to consider the electron inertia in the MHD dynamical equation. Furthermore, we include collisional effects, such as viscosity, and finite resistivity of the plasma (in the behavior of the temperature, we will include the thermal conductivity too). Thus, we deal with the following system of dynamical equations. The first is the generalized Ohm law, obtained from the balance of the forces acting on the electrons, \ie
\begin{align}
\partial_t \vec{J} + \nabla (\vec{J}\cdot\vec{v} + \vec{v}\cdot\vec{J} ) = \frac{n_e e^2}{m_e} (\vec{E} + \vec{v}\!\times\!\vec{B}) + {\eta}_{B} \vec{J} \label{13}\;,\\
\eta_B = \frac{1}{\sigma_B} \equiv \frac{m_e \nu_{ie}}{n_e e^2}\;,
\end{align}
where $n_e$ is the electron number density, $\nu_{ie}$ denotes the ion-electron collision frequency, $\eta_B$ is the resistivity coefficient ($e$ and $m_e$ being the electron charge and mass, respectively), and $\vec{v}$ is the velocity field. In what follows, it will be taken to be purely azimuthal, \ie $\vec{v}=\omega r\hat{e}_\phi$, where $\omega$ denotes the differential angular velocity of the disk.

Then we have the basic law for mass conservation, \ie the continuity equation
\begin{equation}
\partial_t \rho + \rho (\nabla \cdot \vec{v}) = 0\;,
\label{continuity}
\end{equation}
where $\rho$ is the mass density. It is worth noting that, for a purely azimuthal velocity field, from \eref{continuity} we immediately get $\partial_t \rho\equiv0$ and $\nabla\cdot\vec{v}\equiv0$.

The third dynamical equation is the momentum balance in a compressible plasma (\emph{de facto} the MHD extension of the Navier-Stokes equation, including the Lorentz force), \ie
\begin{multline}
\rho [\partial_t \vec{v} + (\vec{v}\cdot\nabla) \vec{v}]=-\nabla p +\\
+\frac{1}{c}\vec{J}\times\vec{B} + \eta_V  \nabla^2 \vec{v}+(\eta_V/3+\iota_V)\nabla(\nabla\cdot\vec{v})\;,
\label{euler}
\end{multline}
where $\eta_V$ denotes the shear viscosity coefficient, $\iota_V$ is the second viscosity coefficient, and $p$ is the thermostatic pressure. We stress that the last term of this equation identically vanishes for the purely azimuthal velocity field at the ground of our analysis, as mentioned above.

It is easy to recognize that \erefs{13} and \reff{euler} have the following non-zero azimuthal components:
\begin{align}
\partial_t \mff = \frac{c^2}{4\pi} \left( \frac{m_e}{n e^2} \partial_t \nabla^2 \mff + \eta_B \nabla^2 \mff \right)
\label{sistema1}\;,\\
\partial_t \mff - \frac{\eta_V}{\rho} \left( \nabla^2 \mff \right) = 0
\label{sistema2}\;.
\end{align}
\eref{sistema2} holds when the corotation condition $\omega=\omega({\mff})$ is assumed \cite{9}. Indeed, when the magnetic field is purely toroidal, we can always require that the azimuthal component $\nabla\times\vec{E}=0$, \ie $\omega=\omega(\mff)$ even in the non-stationary case. Otherwise, for a generic $\omega$, a non-stationary azimuthal component of the magnetic field could be generated: The corotation condition is no longer ensured by a theorem, but it still survives as a particular solution of the non-stationary induction equation. In the vertical and radial directions, \erefs{13} and \reff{euler} result in
\begin{align}
0= \partial_z p + \rho{\omega_k}^2 z - \frac{1}{4\pi r}\partial_z \mff  \tilde{\Delta} \mff
\label{v}\;,\\
-\rho \omega^2 r = -\partial_r p - \rho \omega_k^2 r +\qquad\qquad\qquad\qquad\qquad\nonumber\\
- \frac{1}{4\pi r} \partial_r \mff \Big (\partial_r \Big( \frac{1}{r} \partial_r \mff \Big) + \frac{1}{r} {\partial_z}^2 \mff \Big)\;,
\label{R}
\end{align}
respectively, where $\tilde{\Delta}=\p_r(r^{-1}\p_r)+r^{-1}\p^2_z$.

As previously stressed, the investigation of the evolution of the magnetic flux surface is performed by means of a perturbative approach. Thus, the density and pressure functions are split into two terms around the fiducial radius $R_0$ namely $\rho=\rho_{0}+\rho_{1}$ and $p = p_0 + p_1$. Accounting for the local character of our analysis and the small-scale structure of the perturbation $\mff_1$, the approximation
\begin{equation}
\Big( \partial_r \Big(  \frac{1}{r} \partial_r \mff_1 \Big) + \frac{1}{r} {\partial_z}^2 \mff_1 \Big) \simeq \frac{1}{R_0} \Delta \mff_1 \;,
\label{laplaciano}
\end{equation}
holds, where $\Delta \mff_1 \equiv\partial_r^2 \mff_1 + \partial_z^2 \mff_1$.

Finally, we observe that the background we are perturbing corresponds to a purely differentially rotating disk (\ie $\omega= \omega_0 (\mff_0 (R_0))$) which is embedded in the steady vacuum magnetic field of the central object described by $\mff_0 = \mff_0 (R_0)$ (we are neglecting the plasma backreaction on the background). The perturbation quantities are regarded as varying on small spatial scales. Thus, their gradients can be relevant, especially those of second order which dominate and provide the current density flowing in the disk (which is regarded as negligible on the background). Concerning \erefs{sistema1} and \reff{sistema2}, they hold for the perturbed function $\mff_1$, as well as for $\mff$. This is due to the stationarity of $\mff_0$ and the small scale of variation of $\mff_1$, such that $\nabla^2\mff_1\gg \nabla^2\mff_0$. Therefore, we can rewrite \erefs{sistema1} and \reff{sistema2} as
\begin{align}
\partial_t \mff_1 - \frac{c^2}{4\pi} \left( \frac{m_e}{n e^2} \partial_t \Delta \mff_1 + \eta_B \Delta \mff_1 \right)=0\;,
\label{unos}\\
\partial_t \mff_1 - \frac{\eta_V}{\rho} \left( \Delta \mff_1 \right) = 0
\label{dues} \;.
\end{align}
Below we analyze the obtained dynamical system in two different regimes: Linear (when the backreaction magnetic field is small) and non-linear (when the backreaction magnetic field is comparable to the background one). Since \erefs{unos} and \reff{dues} are intrinsically linear, the crucial difference between the linear and non-linear regimes will consist in the specific form acquired by the perturbed form of \erefs{v} and \reff{R}.

\section{Linear regime}
In the present scheme, the mass density remains a stationary variable because its behavior is governed by the continuity law \reff{continuity}, which provides $\partial_t \rho_1 = 0$ (\ie $\rho = \rho_0(R_0,z)$). Thus, \erefs{sistema1} and \reff{sistema2} can be split to describe the spatial and the temporal behavior of the magnetic flux surface,
\begin{align}
\Delta\mff= \frac{\nu_{ie}\rho}{\eta_V}(\prm-1)\mff\;,\label{spazio}\\
\partial_t\mff=\nu_{ie}(\prm-1)\mff\;,\label{tempo}
\end{align}
where the MPN has been introduced as follows:
\begin{equation}
\prm \equiv \frac{4 \pi \eta_V}{c^2 \rho \eta_B}\;.
\end{equation}
Clearly, the value of $\prm$ influences critically the form of $\mff$ and the solution of \eref{tempo} is
\begin{equation}
\mff(r,z^2,R_0,t)= \bar{\mff}(r,z^2,R_0)e^{\nu_{ie}(\prm-1)t}\;.
\label{solT}
\end{equation}
If $\prm>1 $ or $\prm<1$, we clearly deal with two different regimes corresponding to a growth or a damping of the flux function.

Meanwhile, \eref{spazio} does not admit an analytical general solution. Let us now assume a separable form for the function $\bar{\mff}$, \ie
$$\bar{\mff}(r,z^2,R_0)=N(r,R_0)F(z^2)\;$$
Restricting our analysis to close to the equatorial plane, so that $z/H\ll1 $ ($H$ being half the depth of the disk), and defining the normalized density $D$ as
\begin{equation}\label{dexp}
\frac{\rho(z)}{\rho(z=0)} = D(z^2) = e^{-z^2/H^2} \simeq \Big( 1 - \frac{z^2}{H^2} \Big)\;,
\end{equation}
we finally get the following solution, strictly valid for the case $\prm<1$:
\begin{equation}
\bar{\mff}(r,z^2,R_0) = \bar{\mff}_0^0 \sin[k_2 (r-R_0)] e^{-z^2/\Delta^2}\;.
\label{Psi}
\end{equation}
Here we have introduced the parameters
\begin{align}
\Delta^2 = \frac{2H}{\sqrt{-k_1}}\;,\\
k_1= \frac{\nu_{ie}}{2\alpha v_s H/3} (\prm -1)\;,\\%p
k_2 = \sqrt{-k_1 \Big( 1- \frac{1}{\sqrt{-k_1}H} \Big)}\;.\label{k2}
\end{align}
We stress how we have adopted the standard Shakura expression for the viscosity coefficient, \ie
$$ \eta_V \equiv  \frac{2}{3} \alpha H v_s \rho_0(z=0)\;,$$
where $v_s$ is the background plasma sound velocity and $\alpha$ is a dimensionless parameter.

Thus, in this specific case, a magnetic structure has been found which is periodic in the radial direction, with a temporal damping like in \citer{Petitta2013}. In fact, as shown in \citers{2,1}, under the hypotheses considered, the radial and vertical Navier-Stokes equation components \reff{v} and \reff{R} reduce, in the linear regime $|\partial_r \mff_0|\gg|\partial_r \mff_1 |$, to the single (radial) one:
\begin{equation}
\Delta \mff_1 = - k_0^2 \mff_1\;,\qquad
k_0^2 \equiv \frac{\omega_K^2}{v_A} \;,
\label{nuova}
\end{equation}
where $v_A^2=\partial_r\mff_0^2/(4\pi R_0^2\rho_0)$ is the background Alfv\'en velocity and $\omega_K\equiv\omega_0(\mff_0)$ denotes the Keplerian angular velocity.

It is possible to find a relation between the MPN and the typical $\beta$ parameter of the plasma. By comparing \eref{nuova} with \eref{spazio}, we arrive ot the following identification:
\begin{equation}
k_0^2 = \frac{\nu_{ie} \rho_0}{\eta_V} (1 - \prm)\;.
\end{equation}
Adopting again the Shakura prescription for the $\eta_V$ coefficient and recalling the definition of the classic plasma parameter $\beta$, 
\begin{equation}
\beta= \frac{4 \pi p}{B^2} = \frac{1}{3} H^2 k_0^2 \equiv 1/(3\epsilon_z^2)\;,
\end{equation}
we can easily obtain
\begin{equation}
2 \alpha \omega_K \beta = \nu_{ie} (1 - \prm)\;.
\label{legame}
\end{equation}
Now, using the condition of reality of the root in \eref{k2}, we obtain
\begin{equation}\label{beta25}
\beta>0.25\;,
\end{equation}
which is a restrictive condition for the existence of this periodic structure for the magnetic flux surface.

We now stress how, in the case $\prm>1$, the expression \reff{solT} is associated with an exponential growth of the magnetic flux function. Thus, this regime corresponds to an unstable behavior of the system. However, it is important to stress that \eref{spazio} would provide an intrinsic linear differential problem for the function $\mff_1$. It easy to realize how such an equation would be incompatible with the linear limit \reff{nuova} of the radial configurational equation (since there the sign in the coefficient of the right-hand side is necessarily negative). As a consequence, the unstable behavior, associated with the range of values $\prm>1$, can only survive in the fully non-linear regime, \ie
\begin{equation}
|\partial_r \mff_1| \sim |\partial_r \mff_0 |\;,
\end{equation}
when \eref{nuova} does not hold and it is replaced by a non-linear problem. In this limit, we also observe that the radial dependence of $\mff_1$ changes with respect to the crystalline structure, although remaining a small-scale configuration.

\section{Non-linear regime}

We now address the analysis of the full set of dynamical equations in the non-linear regime where the backreaction magnetic field is comparable to or greater than the background one $|\partial_r \mff_1| \geqslant |\partial_r \mff_0|$. The dimensionless first-order perturbed system reads
\begin{subequations}\label{r}
\begin{align}
\partial_{u^2}\hat{P} + \epsilon_z \hat{D} + 2\Delta_{\epsilon_z}Y \partial_{u^2} Y= 0 \;,\\
\partial_x\hat{P}/2 + (\bar{D} +\hat{D}/\beta) Y + \Delta_{\epsilon_z}Y (1 + \partial_x Y) = 0\;,\\
\partial_{\bar{t}} Y = \gamma Y\;,\label{Y1}\\ 
\Delta_{\epsilon_z} Y = \gamma \bar{D}(u^2) Y \label{Y2}\;,
\end{align}
\end{subequations}
where we have introduced the notation
\begin{align}
Y=\frac{k_0 \mff_1}{\partial_{R_0} \mff_0}\;,\quad x=k_0 r\;, \quad u=\frac{z}{\sqrt{H/k_0}}\;,\nonumber\\
\bar{t}=\frac{2\alpha k_0 v_s}{3\epsilon_z}\;t\;,\qquad\gamma=\frac{3\nu_{ie}\epsilon_z(\prm-1)}{2\alpha k_0 v_s}\;,\nonumber\\
D= \bar{D} + \hat{D}\;,\quad P=\bar{P} + \hat{P}\;,\quad\Delta_{\epsilon_z}\equiv \p^2_x+\epsilon_z\p^2_u\;,
\nonumber
\end{align}
where $D$ is defined in \eref{dexp}, while $P=p/p(z=0)$. Expressions marked with an overbar and a circumflex denote background and perturbation quantities, respectively.

We now observe that \eref{Y1} admits the solution
\begin{equation}
Y(\bar{t},x,u^2) = Y_0(x,u^2)e^{\gamma \bar{t}}\;,
\label{solx1}
\end{equation}
which, substituted in \eref{Y2}, provides the fundamental configurational equation
\begin{equation}
\Delta_{\epsilon_z}Y_0=\gamma(1-\epsilon_z u^2)Y_0\;.
\label{solx2}
\end{equation}
It is easy to check that this equation admits the following solution
\begin{equation}
Y_0=A\; \textrm{Re}\Big[\exp\Big[{x\sqrt{\gamma+\epsilon_z\sqrt{-\gamma}}}\;-u^2\sqrt{-\gamma}\,/2\Big]\Big]\;, 
\label{solfin}
\end{equation}
where the constant amplitude $A$ must be fixed by the initial condition on the real plasma disk.

As it is clearly illustrated by the limit $\epsilon_z\to0$ (\ie the limit of large $\beta$ values, typical of astrophysical regimes), when $\gamma$ (\ie $\prm-1$) is negative, the profile is damped in time and with the vertical height, while it radially oscillates (damped crystalline structure). Otherwise, when $\gamma$ (\ie $\prm-1$) is greater than zero, the configuration takes the morphology of an instability (it growths in time), oscillates in the vertical coordinate, and growths radially too (non-linear unstable regime). In the case of a non-negligible value of $\epsilon_z$, but still small, the situation remains the same, but for $\prm>1$, the radial dependence acquires a small oscillating component in addition to the exponential growth. Finally, we note that when $\gamma$ passes from negative to positive values, we go from from trigonometric functions (intrinsically bounded) to hyperbolic trigonometric functions (in principle divergent). However, their behavior remains valid only near the fiducial radius and therefore they never really diverge.

\begin{figure}[ht!]
\includegraphics[scale=0.6]{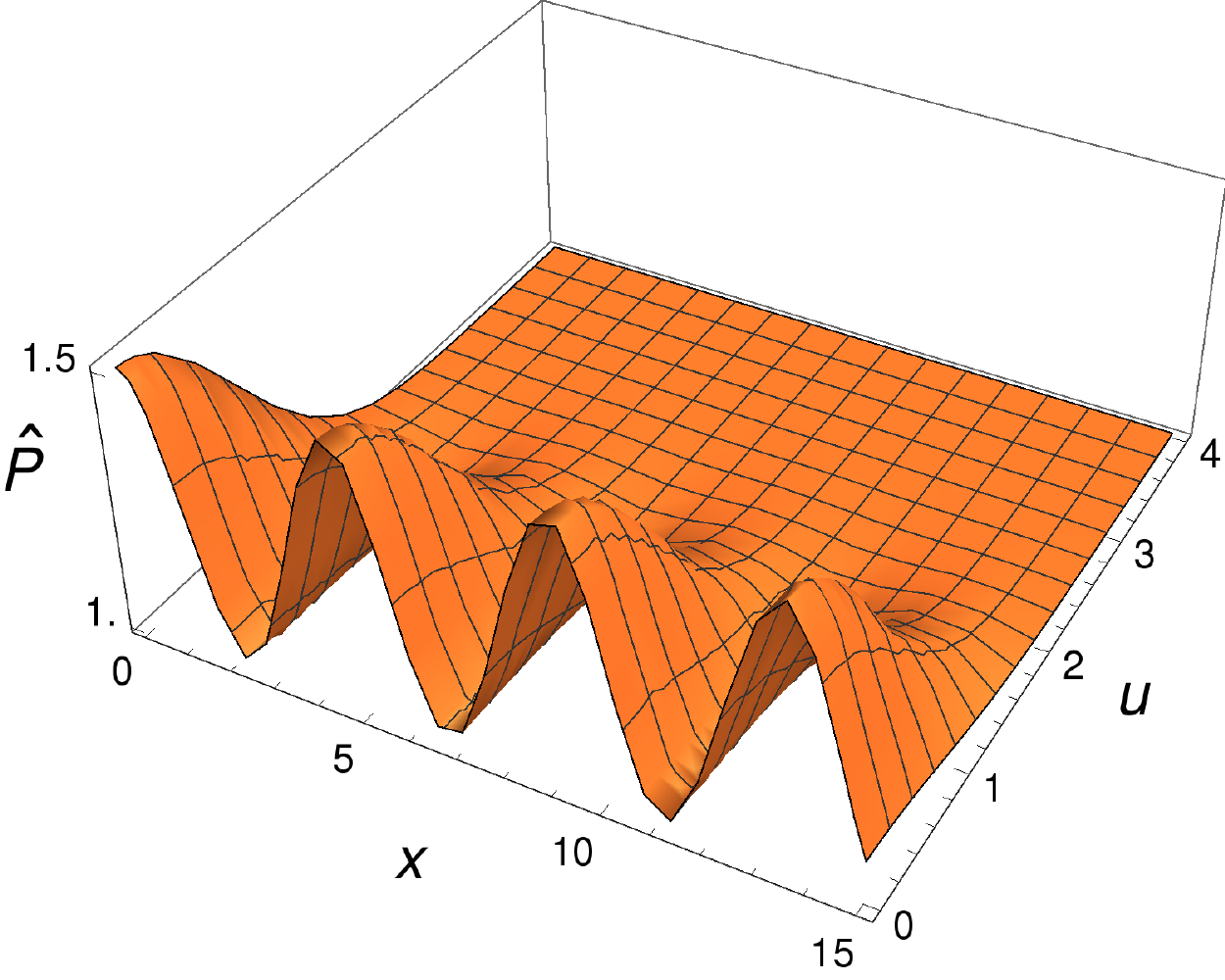}\\
\includegraphics[scale=0.6]{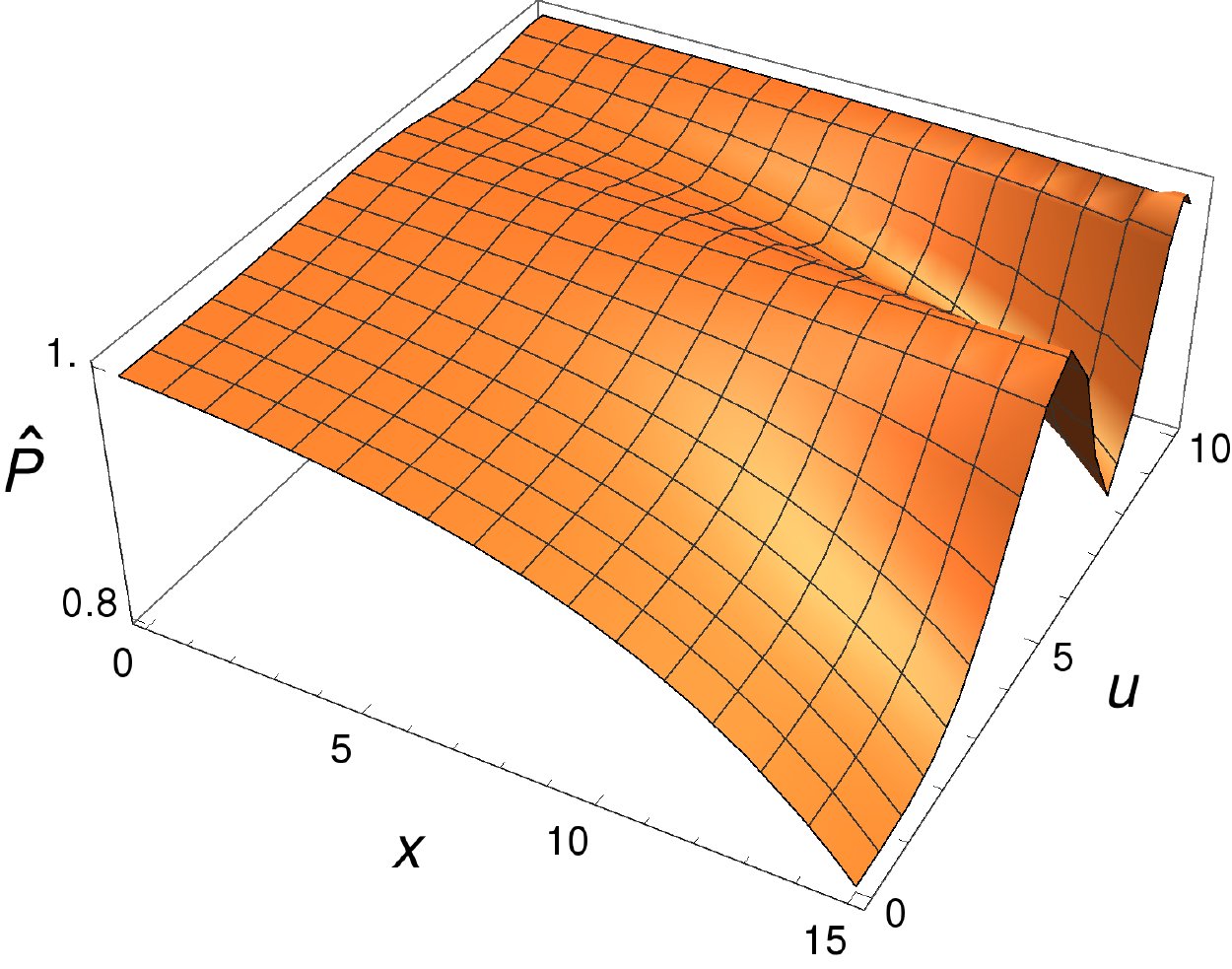}
\caption{Plot of the dimensionless pressure of \eref{press} for $\bar{t}=0$, $\Pi=1$, $\epsilon_z=0.001$, $A=1$ and $\gamma=-0.5$ (top) and $\gamma=0.01$ (bottom).\label{fig1}}
\end{figure}

Let us now look for a general solution of the non-linear system above. For the stationary form of the density previously discussed, we examine the regime where $|Y|\gg1$ and both $\hat{D}$ and the linear terms in the first two equations of the system \reff{r} are negligible. In this way, only the two equations
\begin{align}
\partial_{u^2}\hat{P} + 2 \;\gamma\, \bar{D}  \, Y \, \partial_{u^2} Y= 0\;,\\
\partial_x\hat{P} +  2 \gamma\,\bar{D} \, Y \, \partial_{x} Y= 0 \;,	
\end{align}
survive and we get the solution
\begin{equation}\label{press}
\hat{P} = [\Pi(R_0,z_0) - \gamma \bar{D} Y_0^2 ] e^{2\gamma\bar{t}}\;.
\end{equation}
In this expression, $\Pi$ is an integration constant and we stress how the perturbed pressure depends quadratically on the function $Y$. Thus, for $\prm<1$, it exhibits a periodic structure just like the magnetic flux function, as sketched in \fref{fig1}.

It is worth noting that such behavior of the pressure holds for both $\prm<1$ and $\prm>1$, although the latter exists only in this non-linear regime and it is not associated with a crystalline structure, while the former is present, as shown in the preceding Section, even for a weak backreaction of the plasma (and it always corresponds to a radial oscillation). Actually, \erefs{Y1} and \reff{Y2} are intrinsically linear and therefore hold for any intensity of the backreaction. The present analysis demonstrates that, for $\prm>1$, a non-linear instability exists and it is described by an exponential growth of the perturbed magnetic flux function and of the corresponding thermodynamic pressure contribution. Indeed, in the non-linear limit, a bifurcation takes place: If the resistivity dominates over the viscosity contribution ($\prm<1$) the crystalline structure is damped, while in the opposite regime ($\prm>1$) a new non-linear regime is present.

Regarding the regime in which the crystalline structure is damped, we stress that the present analysis extends the study in \citer{Petitta2013}, valid for $\prm=1$, to the whole region $0<\prm<1$. This allows a much longer duration of such transient processes. This is an interesting issue because it permits one to apply the present mechanism to a wider class of astrophysical process, like the cataclysmic variables. 

\subsection{Role of temperature}
We now briefly investigate the behavior of the disk plasma temperature, during the evolution of the structures outlined above, both in the presence of damping and when the non-linear instability is triggered. First of all, it is worth expressing the dependence of the model parameters on the temperature, namely, we have
\begin{align}
\nu_{ie} = \frac{4}{3} e^4 n_e \frac{\sqrt{2 \pi}}{\sqrt{m_e}} \frac{1}{T^{3/2}} \ln(\Lambda_e)\;,\\
\eta_B = \frac{m_e \nu_{ie}}{n e^2} \sim T^{-3/2}\;,\\
\eta_V = \frac{m_i n \, v_s^2}{\nu_{ie}} \sim T^{5/2}\;,\\
\prm = \frac{4\pi \eta_V}{c^2 \rho \eta_B} \sim \frac{T^4}{n_e \ln(\Lambda_e)}\;,
\end{align}
where $\ln(\Lambda_e)$ denotes the Coulomb logarithm. In particular, we stress how our critical parameters $\nu_{ie}$ and $\prm$ have the opposite behavior in terms of the temperature: The former decreases with $T$ while the latter increases.

It is well known that Coulombian collisions in a plasma weakly affect its internal energy with respect to the ideal gas expression (this can be assumed true also in the presence of effective dissipation due to turbulence). Thus, we are led to infer that the perfect gas equation of state, here postulated for the adiabatic background, remains valid at the first order of perturbation. However, we have to emphasize that, in the present case, both the divergence of the velocity field and the advective operator identically vanish. As a consequence, the evolution of pressure and temperature (here the mass density is necessarily constant in time) must nonetheless be governed by the same dissipation contribution.

Regarding the non-linear case above, we can obtain, using the ideal gas equation of state $p=k_B T \rho/m_i$ (where $m_i$ is the ion mass and $k_B$ the Boltzmann constant), a relation between $P_1$ and $T_1$. In fact, if we split the temperature into the background contribution $T_0$ and the perturbed term $T_1$, we obtain
$$ p= \frac{k_B T}{m_i} \rho \;\;\Rightarrow\;\; P_1 =\frac{k_B \rho_0}{m_i} T_1\;, $$  
where $\rho_0$ is constant in time. Therefore, the perturbed temperature acquires the same behavior of the pressure, namely,
\begin{equation}
T_1 \sim (\prm-1)\mff_1^2\;.
\label{propto}
\end{equation}
Thus, requiring a quasi-ideal behavior of the disk plasma, we realize that the temperature must evolve both with time and with the function $\mff_1$ itself (at least in the perturbed scheme).

In general, the equation governing the temperature evolution contains all Joule, viscous, and finite electron inertia contributions. However, when viscosity is present in the system, also thermal conductivity must be accounted for and it provides the typical diffusion term of the thermal energy. If we postulate that such a term dominates the temperature dynamical equation, we get 
\begin{equation}
\frac{3}{2}\frac{\rho_0}{m_i}\, k_B \, \partial_t T = \kappa_T \Delta T\;,
\label{T}
\end{equation}
where $\kappa_T$ is the thermal conductivity coefficient. Immediately, \eref{T} reverts to \eref{sistema2} and leads to a relation between the temperature and magnetic flux surface, \ie $T=T(\mff)$. Considering \eref{propto}, it is possible to rewrite \eref{T} as a function of $\mff_1$. According to the gradient hierarchy already introduced in the perturbation scheme above, we can neglect the quadratic gradient of $\mff_1$, \ie the following condition holds:
$$ | ( \nabla \mff_1  )^2 | \ll | \Delta \mff_1 |\;.$$
Thus, we easily obtain \eref{sistema2} if the following constraint for the thermal conductivity is valid:
$$ \kappa_T= \frac{3}{2}  k_B \frac{\eta_V}{m_i}\;.$$

Furthermore, considering $T=T(\mff)$ and splitting the different orders, we then get
\begin{align}
T(\mff)=T_0(\mff_0)+\frac{\partial T}{\partial \mff} \mff_1 +
\frac{1}{2}\frac{\partial^2 T}{\partial {\mff}^2} {\mff_1}^2\simeq\nonumber\\
\simeq T_0(\mff_0)+\frac{1}{2}\frac{\partial^2 T}{\partial {\mff}^2} {\mff_1}^2\;,
\label{sviluppo}
\end{align}
where we accounted for \eref{propto}, which implies that
$$\frac{\partial T}{\partial \mff}=0\;.$$
This means that $\mff_0$ is a stationary point for the temperature evolution. In particular, it corresponds to a maximum value in the damped case $\prm<1$ and to a minimum where the non-linear instability takes place for $\prm>1$.

Thus, starting from our guess about the evolution of the temperature as guided by the thermal diffusion only (which appears certainly well-posed for large values of $T_0$ and $\rho_0$ in the kinetic limit or for large value of the viscosity coefficient), we arrive at the construction of a consistent behavior for all the system variables, able to preserve the quasi-ideal feature of the plasma disk. This is in agreement with a physical prediction of the behavior of the temperature in the two regimes $\prm>1$ and $\prm<1$: In the former case, the plasma temperature starts to increase from a minimum value as an effect of the non-linear instability, while in the latter it is damped by the dissipation. Clearly, the scenario traced above is not unique, due to the large number of different regimes able to take place in different domains of the model parameters.

\section{Estimate of damping time}
Let us now investigate the temporal duration of the structures described above. In order to be observed, the micro-structures must exist beyond the dynamical time scale $1/\omega_K$, which is the time needed for the vertical hydrostatic equilibrium to be established, and we assume that it is preserved. Thus, the condition
$$\tau\omega_K\gg1\;,$$
must hold, where $\tau$ denotes the life-time of the micro-structures, in order for the model to be consistent and predictive for astrophysical processes.

\begin{figure}[h]
\includegraphics[scale=0.55]{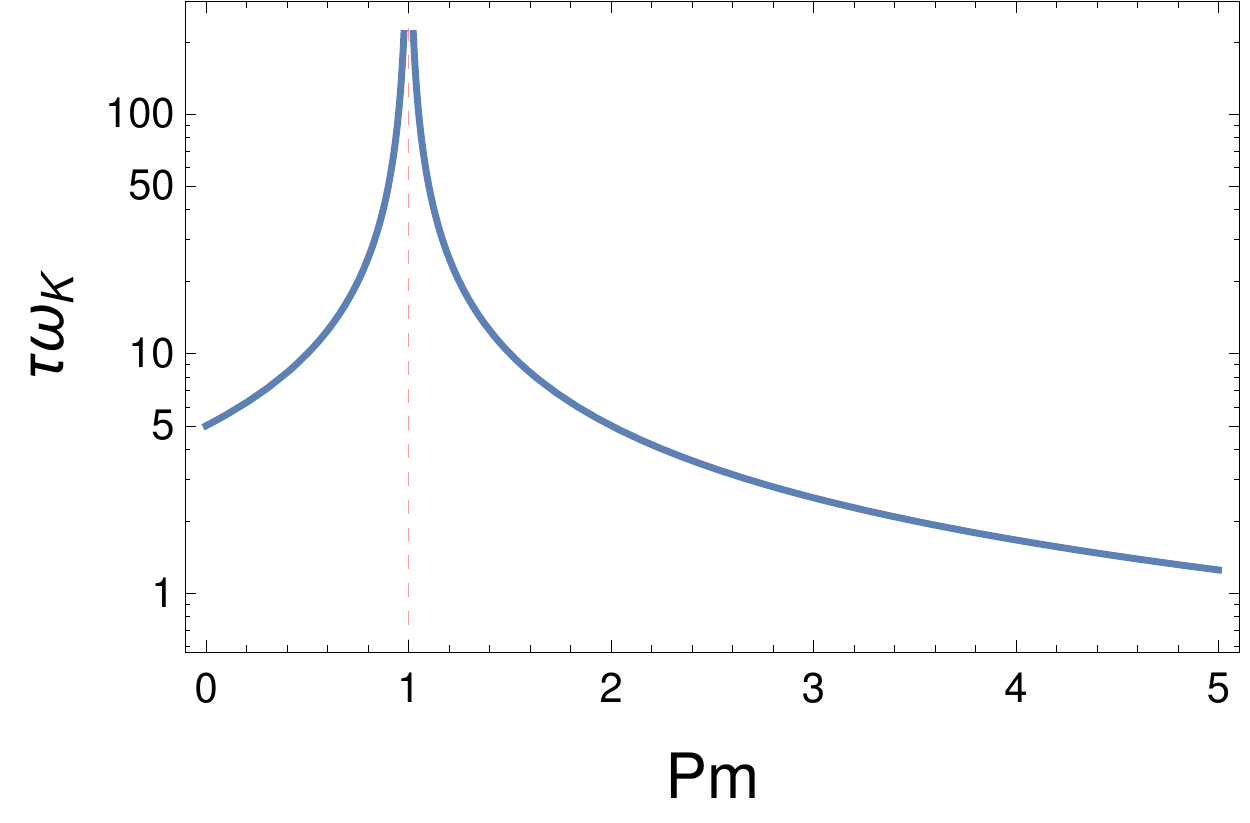}
\caption{Plot of $\tau\omega_K(\prm)$ in \eref{nmnabhasdb}. To illustrate the behavior of such a quantity, we set $H^2\nu_{ie}\rho_0/(3\eta_V)=1$. The dashed red line represents the asymptote for $\prm=1$.\label{fig2}}
\end{figure}

This ratio can be explicitly found in terms of the three variables ($T_0$, $\rho_0$, and $R_0$):
\begin{align}\label{hkjhkfjdhf}
\tau\omega_K \sim \frac{T_0^{3/2}R_0^{-3/2}}{\rho_0(m_i T_0^4 /\rho_0-1)}\;,
\end{align}
where all the parameters depend on the physical features of the stellar object. In order to estimate $\tau\omega_K$, we observe that it can be written, by means of \eref{legame}, as
\begin{align}\label{nmnabhasdb}
\tau\omega_K = 1/|2\alpha\beta|\;.
\end{align}
In the case $\prm>1$, by estimating \eref{hkjhkfjdhf} for quasi-ideal kinetic values of the parameters we get $\tau\omega_K\lesssim1$. However, $\prm$ can receive contributions by effective dissipation due to turbulence. In this case, for $\prm$ slightly greater than one, the time scale of the non-linear instability can be very large, as depicted in \fref{fig2}. Meanwhile, when $\prm<1$, accounting for the convention of $0.01<\alpha<0.1$, and $\beta>0.25$ (see \eref{beta25}), we can conclude that, in this case, $\tau\omega_K$ is always greater than one and therefore the perturbed plasma configurations discussed above survive for a sufficient long time to get astrophysical meaning since they could be involved in the mechanism of angular momentum transport.

This constitutes a significant upgrading of the analysis in \citer{Petitta2013}, since the duration of the transient process is enhanced. Indeed, in the present model, for $\prm$ slightly greater or less than one ($\prm=1$ is the case studied in \citer{Petitta2013}), the time scale of the micro-structures can be much greater (see \fref{fig2}) than $1/\omega_K$ and they becomes of interest for a wider class of astrophysical phenomena.

Instead, in the case of $\prm >1$, the increase of $\prm$ decreases the time scale and so does the possibility for the structures to exist. The very small value of the characteristic time in the case $\prm>1$ for kinetic values of the parameters is clearly consistent with the emergence of an instability which is just the trigger of an incoming process. We conclude by emphasizing how the growth rate of such non-linear instability can be much greater than all other linear instabilities present in the disk (for instance, the fundamental MRI one) whose characteristic time is of order $1/\omega_K$.

\section{Conclusions}
In this work we analyzed the linear and non-linear behavior of a thin disk configuration, whose background profile is a purely differentially rotating plasma, embedded in the gravitational field of the central object. The triggered perturbations preserve both the co-rotation condition and the negligibility of poloidal velocity components.

In the present model we included both ideal effects (like the finite electron inertia) and collisional corrections to MHD, in particular finite plasma electric conductivity, viscosity, and thermal conductivity. In this respect, two main different regimes have been identified: \emph{(i)} The limit in which the resistivity of the plasma dominates the viscosity (MPN number less than one), where both the linear and the non-linear perturbation evolution can be addressed, and \emph{(ii)} the opposite case of dominating viscosity (MPN greater than one), where the non-linear perturbation dynamics is only available.

With respect to the first regime, when the crystalline profile of the disk is damped by the collisional effects, the main merit of the present analysis has been to extend the results obtained in \citer{Petitta2013} (valid only for MPN exactly equal to one) toward a wider class of behaviors. As discussed in \citer{Bal98}, the standard model for accretion disk relies on very small values of MPN, available in the proposed scenario. In particular, the duration (the mean life-time) of the crystalline structure is significantly enhanced in the present model, allowing its implementation to describe a wider class of astrophysical transients. In other words, we upgrade the previous analysis in \citers{2,Petitta2013}, demonstrating how the radial oscillation of the magnetic flux function, due to the plasma backreaction and originally outlined in \citer{1}, is significantly affected by collisional effects. However, such a damping allows the micro-structures to survive for a sufficiently long time to be correlated with transient astrophysical phenomena, like the jet formation or the dynamics of cataclysmic variables.

The regime dominated by the viscosity offers the most intriguing feature emerging from the present analysis, \ie the existence of a non-linear instability of the system. This is characterized by very high growth rates (at least for $\prm$ significantly different from one) and is able to enhance the crystalline profile of the disk toward new plasma configurations, presumably associated with saturation processes of such an instability. Again, the rapid evolution of this new regime suggests that it could concern the triggering of physical processes across the thin disk configuration, emerging from a change of pre-existing conditions of the plasma. In particular, we observe that such a limit $\prm> 1$ corresponds to the real kinematic properties of the plasma which is, in many accretion disk regions, quasi-ideal (see \citers{Bal98,Petitta2013}). Thus, we are lead to infer that the new non-linear instability we trace here is triggered by a significant suppression of the disk turbulence, responsible for the effective value $\prm\ll1$, like in the $\alpha$ models \cite{Bal98}; indeed, in the absence of turbulence, the viscosity and resistivity of the disk take their quasi-ideal value, which corresponds to $\prm\gg1$ \cite{Petitta2013}.

A possible scenario in which the transition from the damped to the unstable regime is non-linearly viable could correspond to a rapid cooling of the disk with the associated suppression of the MRI and of the corresponding turbulence. In this respect, we observe that the temperature is indeed suppressed in the damping regime of the crystalline structure, suggesting the following intriguing paradigm: If the disk backreaction is of small scale, a crystalline configuration of the disk can be achieved, but its evolution is strongly affected by the effective viscosity and resistivity present in the disk so that its profile is damped, together with the disk temperature. This cooling of the disk suppresses MRI and then turbulence, restoring the quasi-ideal character of the plasma which, in the non-linear regimes, induces the triggering of a new instability. However, the validation of such a paradigm requires that some non-trivial questions must be addressed, including \emph{(i)} the clarification of the real process (maybe external to the disk physics, like a sound or a gravitational wave impacting it), which is able to determine the existence of the crystalline morphology of the radial profile, and \emph{(ii)} the demonstration that the cooling phase of the disk takes place in the non-linear regime, where the instability can be triggered. Nonetheless, the main merit of the present analysis consists in tracing a new possible scenario for accretion disk non-linear instability.

\end{document}